\documentstyle[aps,floats,epsf,twocolumn]{revtex}

\begin{document} \draft

\title{Flux tubes, visons, and vortices in spin-charge separated 
superconductors}

\author{M. Franz$^*$ and Z. Te\v{s}anovi\'c$^\dagger$}
\address{
Institute for Theoretical Physics, University of California,
Santa Barbara, CA 93106
\\ {\rm(\today)}
}
%\maketitle
%
%\begin{abstract}
\address{~
\parbox{14cm}{\rm
\medskip
The idea of spin-charge separation in cuprate superconductors has been 
recently energized by Senthil and Fisher who formulated a Z$_2$ 
gauge theory and, within its context, 
proposed a ``vison detection''  experiment as a  test for  topological
order in a sample with multiply connected geometry. Here we show that the same
experiment can be performed to test for the spin-charge separation
in U(1) [but not in SU(2)]
theory and argue that vortex core spectroscopy can in fact distinguish 
between the different symmetries of the fictitious gauge field.
}}
%\end{abstract}
\maketitle

%\pacs{74.60.-w,74.60.Ec,74.72.-h}

%
\narrowtext

While there exist number of compelling reasons to think that electron could
be fractionalized in high-$T_c$ cuprate superconductors\cite{anderson1}, 
there is thus far
no {\em direct} experimental evidence for this effect in dimensions $D$ higher 
than 1. Recently, Senthil and Fisher (SF) proposed an experiment
\cite{senthil1} to 
directly test for electron fractionalization in $D=2,3$ in strongly underdoped
cuprate samples with multiply connected geometries. SF framed their proposal
in the context of the  Z$_2$ gauge theory\cite{senthil2}. 
The purpose of this note is to point out that SF experiment constitutes
a general test for spin-charge separation, and, as long as the charge
carrying boson (be it holon or chargon) condenses without pairing, the outcome
will be independent of the symmetry of the fictitious gauge field present in
the theory. Thus, if the spin-charge separation takes place in cuprates, 
the outcome of the SF experiment will be the same for U(1) and Z$_2$ 
theories. As pointed out by Lee and Wen\cite{lee2}, it will not work in
the SU(2) theory.
We furthermore point out that SF experiment is closely tied to the
ongoing debate on the structure of a magnetic vortex in spin-charge separated 
superconductors\cite{lee2,sachdev1,nagaosa1,franz1,han1} and propose that 
probing the excitation spectrum in the core 
of a vortex as a function of doping can shed light on the symmetry of the 
fictitious gauge field.

The essence of the SF experiment\cite{senthil1} lies in
the realization that a singly
quantized vortex carrying $hc/2e$ magnetic flux is a very peculiar object
in the fractionalized superconductor. This is because charge carrying boson 
of the theory (holon or chargon) is assumed to Bose-condense individually,
i.e. without pairing. Such charge-$e$ condensate would normally quantize the
magnetic flux in multiples of $hc/e$. Ordinary superconducting $hc/2e$ vortex  
would cause the condensate wavefunction to acquire a phase $\pi$ on 
encircling the vortex, producing a branch cut.
Branch cut in the macroscopic wavefunction would lead to various
catastrophic consequences (such as infinite currents) and must be 
therefore compensated by a feature in the gauge field. In the
Z$_2$ theory this is done by binding ``vison'', a topological excitation
of the  Z$_2$ gauge field, to the
singly quantized vortex. Vison supplies the missing $\pi$ phase to the
chargon condensate wavefunction but does not alter the spinon condensate 
wavefunction because the latter are assumed to condense in singlet pairs and
therefore the net phase acquired by paired spinons is $2\pi$. 

SF envision trapping
a singly quantized vortex in the hole fabricated in a strongly underdoped
superconductor. Such hole would then necessarily be threaded by a vison. 
When heated above the superconducting $T_c$, the magnetic flux can easily 
escape from the hole but vison remains trapped because it is a gapped 
excitation below the pseudogap temperature $T^*$. Although there is no direct 
way to detect vison, one could deduce its presence by cooling  in zero field
back to the superconducting state, where the vison will bind the magnetic 
flux which can be directly measured.

We now argue that 
the same will happen in the U(1) theory, the main difference being 
that the vison will be replaced by a flux quantum\cite{remark2}
of the U(1) gauge field. 
To illustrate our arguments it is easiest to consider the
effective Ginzburg-Landau (GL) theory
for fractionalized superconductor formulated originally by Sachdev
\cite{sachdev1} and by Nagaosa and Lee\cite{nagaosa1}. The corresponding free
energy reads
\begin{eqnarray}
f_{\rm GL}&=&|(\nabla-2i{\bf a})\Delta|^2 +{r_\Delta}|\Delta|^2 +
{1\over 2}{u_\Delta}|\Delta|^4\nonumber \\
&+&|(\nabla-i{\bf a}-ie{\bf A})b|^2 +r_b|b|^2 +{1\over 2} u_b|b|^4 
 +v|\Delta|^2|b|^2 \nonumber \\
&+& {1\over 8\pi}(\nabla\times{\bf A})^2 +f_{\rm gauge}[{\bf a}],
\label{fgl1}
\end{eqnarray}
where  $\Delta$ and $b$ are spinon pair and holon condensate fields 
respectively, minimally coupled to the electromagnetic field ${\bf A}$ and 
a fictitious U(1) gauge field ${\bf a}$. 
\begin{equation}
f_{\rm gauge}={\sigma\over2}(\nabla\times{\bf a})^2
\label{fgauge}
\end{equation}
is the gauge field stiffness term
which originates from integrating out the high-energy microscopic degrees of 
freedom. It has been argued recently\cite{nayak1} that $\sigma$ is small or 
even zero in realistic models of cuprates. This results in the above GL theory 
being extreme type-I with respect to ${\bf a}$ but extreme type-II with 
respect to ${\bf A}$. In the absence of field the parameters $r$, $u$, and $v$
can be chosen 
such that this GL theory reproduces the standard phase diagram of cuprates
shown in Figure 1. Also, we note that ${\bf a}$ is by construction a
{\em compact} lattice gauge field. Maxwell form of Eq.\ (\ref{fgauge}) is a 
simplified continuum representation of the corresponding lattice expression. 

We first review what is known about the vortex solutions of Eq.\ (\ref{fgl1})
and then we discuss SF experiment in the context of U(1) theory.
As shown in Refs. \cite{nagaosa1,franz1}, in the limit of negligible 
$\sigma$ the free
energy (\ref{fgl1}) allows for two types of singly quantized vortices:
(i) spinon vortex in the overdoped region and (ii) holon vortex in 
the underdoped region [cf. Figure 1]. 
In the spinon vortex the gauge field ${\bf a}$ develops
a net flux such that ${\bf a}\approx -e{\bf A}$ and the singularity occurs
in the spinon field [first term in Eq.\ (\ref{fgl1})] with $\Delta$
vanishing at the vortex center. In the holon vortex $b$ vanishes in the core
and the gauge flux is 
contracted into a $\delta$-function flux tube at the vortex center. This 
fictitious flux
tube is invisible to the spinon condensate and its main purpose is to heal 
the branch cut that would otherwise occur in the
holon order parameter, just as vison does in the Z$_2$ theory.  
In this respect the singly quantized holon vortex in the compact 
U(1) theory is equivalent to the Z$_2$ vortex\cite{remark1}.
\begin{figure}[t]
\epsfxsize=8.5cm
\epsffile{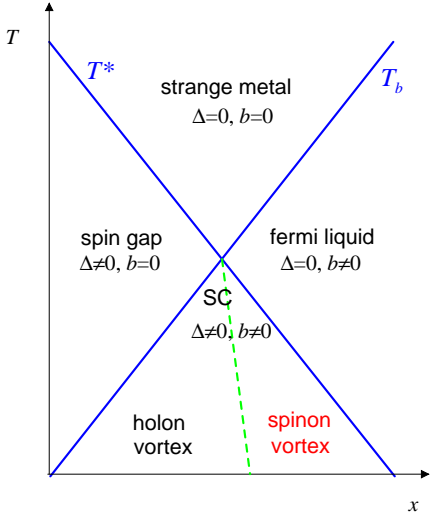}
\caption[]{Schematic phase diagram of a spin-charge separated~ system 
within the U(1) theory in the doping$-$temperature plane, as applied to 
cuprate superconductors.}
\label{fig1}
\end{figure}

The important point to  
notice is that both types of singly quantized vortices in the U(1) theory 
bind a flux $h/2$ (in units where magnetic flux quantum is $hc/2e$) of
the fictitious gauge field ${\bf a}$. Thus, when such vortex is trapped in 
the hole, as envisioned in SF experiment, that hole will be threaded by a
single quantum of the fictitious flux. Now consider removing the external 
field and heating the sample above $T_c$. The magnetic flux will quickly 
escape but, in the strongly underdoped material, the fictitious flux will 
remain trapped below $T^*$. 
To see this, consider that according to the standard U(1) 
phenomenology\cite{lee1} the mean field state just above $T_c$ in the 
underdoped region is characterized by $b=0$ and $|\Delta|>0$. The free energy
(\ref{fgl1}) therefore becomes
\begin{eqnarray}
f'_{\rm GL}&=&|(\nabla-2i{\bf a})\Delta|^2 +{r_\Delta}|\Delta|^2 +
{1\over 2}{u_\Delta}|\Delta|^4\nonumber \\
&+& {1\over 8\pi}(\nabla\times{\bf A})^2 +
{\sigma\over2}(\nabla\times{\bf a})^2. 
\label{fgl2}
\end{eqnarray}
This expression 
is formally equivalent to the GL theory for a superconducting order 
parameter $\Delta$ minimally coupled to the U(1) gauge field ${\bf a}$.  The
main point is that below the critical temperature for $\Delta$, which is
$T^*$ in this model, free energy (\ref{fgl2}) will exhibit the Meissner
effect with respect to ${\bf a}$. On the mean field level the gauge flux
threading the hole cannot penetrate into the bulk below $T^*$ and remains 
trapped there forever. Going beyond the mean-field considerations the 
fictitious gauge flux can tunnel out of the hole, but sample geometry can
be designed in such a way that it remains trapped for sufficiently long times.

The real electromagnetic field ${\bf A}$, on the other hand, is completely 
decoupled from the matter fields. This reflects the obvious fact that
non-superconducting state above $T_c$ cannot exhibit the 
true Meissner effect.

We note that the same argument seemingly could be made for the 
{\em overdoped} region, where above $T_c$ the mean field state is 
characterized by $\Delta=0$ and $|b|>0$. It would appear that the fictitious
flux now remains trapped above $T_c$ by virtue of the Meissner effect caused
by the holon condensate. However, upon closer examination one finds that this 
is not the case because singly condensed holons cannot support
$h/2$ flux quantum in the absence of physical magnetic field. Consequently,
in the U(1) theory the SF effect will occur in the underdoped but not in the 
overdoped region, just as in the Z$_2$ theory.

Having argued that the SF experiment will yield the same outcome irrespective
of the symmetry of the gauge field, we now turn to the differences between
Z$_2$ and U(1) formulations. A qualitative difference occurs inside the vortex
core and can be potentially detected by scanning tunneling spectroscopy (STS).
We emphasize that here we consider a true vortex with the core situated in 
the bulk superconductor, as opposed to the flux trapped in the hole.
As stated above, the singly quantized Z$_2$
vortex (binding a vison) is essentially identical to the singly quantized 
holon vortex in the U(1) theory. However, 
in the U(1) theory there is a transition with increasing doping to the
state in which the spinon vortex becomes energetically favorable\cite{franz1}.
Spectroscopically,
holon vortex should exhibit a pseudogap-like local density of states (LDOS), of
the type currently observed in experiments on  Bi$_2$Sr$_2$CaCu$_2$O$_8$
\cite{renner1,pan1}. Spinon vortex, on the other hand,
should exhibit a more conventional LDOS
with large resonance at the Fermi level, as predicted from $d$-wave BCS 
theory\cite{wang1,franz2}. Transition from holon to spinon vortex as a 
function of doping predicted by U(1) theory should therefore be directly 
observable by STS. In contrast, as argued below, no such transition occurs
within the Z$_2$ theory.

Existence of the spinon vortex in the U(1) theory 
is predicated upon the fact that the fictitious gauge field ${\bf a}$ has the 
same symmetry as the electromagnetic field  ${\bf A}$. Therefore, if it
becomes energetically favorable,  ${\bf a}$
can completely screen the applied magnetic field in the holon term of Eq.\
(\ref{fgl1}) and shift the singularity to the spinon term. Z$_2$ gauge field,
being by definition discrete, cannot do this 
and there can only be one type of a vortex in the Z$_2$ theory. 

In the SU(2) theory the holon condensate kinetic energy has the form
\cite{lee0}
\begin{equation}
|(\nabla +i{\bf a}^{(3)}\hat\tau_3-ie{\bf A})z|^2,
\label{su2}
\end{equation}
where $z=(z_1,z_2)$ is the SU(2) holon doublet, and ${\bf a}^{(3)}$ is the 
component of the gauge field associated with the $\hat\tau_3$ Pauli matrix.
Because of the matrix structure of Eq.\ (\ref{su2}), gauge field 
${\bf a}^{(3)}$ can 
screen magnetic flux seen by one component of $z$, but {\em not both}.
As discussed by Lee and Wen\cite{lee2}, this results in $2\pi$ phase winding 
and suppression in the core of one of the components of $z$. Accordingly,
the vortex  core in SU(2) theory will be in a staggered flux phase.
In the SU(2) theory the staggered flux phase is a gauge equivalent of the 
fermion pairing phase but it is easy to see that ${\bf a}^{(3)}$ does {\em not}
couple to the fermions as a magnetic field. There will therefore be no analog
of the spinon Meissner effect with respect to ${\bf a}^{(3)}$ and the
gauge flux can escape from the hole when the superconducting order is 
supressed.

In summary, we argued that in a spin-charge separated superconductor 
the general outcome of the SF experiment\cite{senthil1} will be the same
for U(1) and Z$_2$ theories. This is a
direct consequence of the fact that charge carrying bosons are assumed to 
Bose-condense individually and the singly quantized magnetic 
vortex must therefore bind additional flux quantum of the fictitious gauge 
field. When such a vortex is trapped in the hole and then removed by heating 
above $T_c$, the fictitious flux cannot 
escape below the pseudogap temperature $T^*$ because it continues to
experience the Meissner effect caused by the spinon pair condensate. 
The trapped flux can then be detected by cooling down below $T_c$ where it 
necessarily binds a quantum of physical magnetic field. 

In the event of positive outcome of the SF experiment it could make sense 
to carry out detailed spectroscopic study of the vortex cores in cuprates 
as a function of doping in order to
establish the symmetry of the fictitious gauge field. We argued that Z$_2$
theory can support only one type of a vortex with pseudogap-type spectrum in 
the core. U(1) theory, on the other hand, predicts a transition from the
holon vortex in the underdoped to the spinon vortex in the overdoped region 
with qualitatively different spectroscopic signatures \cite{franz1}.
In the SU(2) theory of Lee and Wen\cite{lee1} the SF experiment will not 
work. Lee and Wen\cite{lee2} argued for a vortex core in the staggered flux 
phase, and proposed various probes to detect 
its signature. Experimentally there exists evidence for one type of a vortex
in Bi$_2$Sr$_2$CaCu$_2$O$_8$ with pseudogap-type spectrum\cite{renner1,pan1},
but detailed studies of strongly underdoped and overdoped regions have not 
yet been completed. 

The authors are grateful to P. A. Lee and  T. Senthil
for insightful discussions.


\begin{references}
\bibitem{remark2} Here and hereafter we reserve the term ``flux quantum''
for $hc/2e$, i.e.\ the conventional superconducting flux quantum, and the 
corresponding expression $h/2$ for the quantum of fictitious gauge flux.
\bibitem{remark1} The authors are indebted to P. A. Lee for pointing out 
this fact and for correcting their original misunderstanding of the SF
experiment in the SU(2) theory.
\end{references}
\end{document}